\documentclass[twocolumn,preprintnumbers,amsmath,amssymb]{revtex4}
   \usepackage{amssymb}
\usepackage{amsmath}
\usepackage{graphicx}
\usepackage{setspace}

\begin{document}
\title{Networks of strong ties}

\author{Xiaolin Shi}
 \email{shixl@umich.edu}
 \affiliation{Department of Electrical Engineering and Computer Science\\ University of Michigan}
\author{Lada A. Adamic}
 \email{ladamic@umich.edu}
\affiliation{
School of Information\\
University of Michigan}
\author{Martin J. Strauss}
 \email{martinjs@eecs.umich.edu}
 \affiliation{Department of Mathematics\\ Department of Electrical
 Engineering and Computer Science\\ University of
Michigan}

\begin{abstract}
Social networks transmitting covert or sensitive information cannot
use all ties for this purpose. Rather, they can only use a subset of
ties that are strong enough to be ``trusted''. In this paper we
consider transitivity as evidence of strong ties, requiring that
each tie can only be used if the individuals on either end also
share at least one other contact in common. We examine the effect of
removing all non-transitive ties in two real social network data
sets. We observe that although some individuals become disconnected,
a giant connected component remains, with an average shortest path
only slightly longer than that of the original network. We also
evaluate the cost of forming transitive ties by deriving the
conditions for the emergence and the size of the giant component in
a random graph composed entirely of closed triads and the equivalent
Erd\"os-Renyi random graph.
\end{abstract}

\maketitle

\section{INTRODUCTION}
The strength of weak ties is the concept that individuals tend to be
more successful in acquiring information about job opportunities by
contacting individuals that they did not see often---their weak
ties~\cite{granovetter73ties}. The rationale behind this idea is
that close friends tend to have similar information to us because
they share similar interests, profession, or geographical location.
Weak ties on the other hand are between individuals who don't have
much in common, including other contacts, and the information they
have access to will tend be different. A shared contact between two
individuals forms a closed triad (triangle), where all three people
know one another. Strong ties are usually parts of triads because
good friends or close professional contacts of one person will tend
to know one another. In this paper we make the simplifying
assumption that `weak ties' are those that are not part of any
closed triad and we define `strong ties' are the ones that share at
least one other contact in common. In other contexts the strength of
the tie may include measures such as frequency or length of contact,
but for simplicity here we consider only the presence of closed
triads.

While weak ties may be preferred in acquiring job information, one
may be interested in assembling a team or otherwise gathering
information that is distributed in different parts of a social
network using only strong ties. In the case of the Madrid
terrorist bombings on March 11th, 2003, the individuals behind the
attack were able to procure knowledge about making explosive
devices, hashish to trade for explosive materials, and the
explosive material itself using their strong ties. Had they used
weak ties which would have been less reliable, their plot may have
been exposed and their intentions thwarted. Sinister plots are not
the only example of a planning activity that can benefit from
using strong ties to maintain confidentiality. Scientists may wish
to forge collaborations requiring diverse expertise
\cite{guimera2005team}, and in doing so they may wish to keep a
competitive edge by not broadcasting their ideas over weak ties.
Similar situations may arise in the formation of business
alliances, where companies seek to complement their strengths
through mergers, acquisitions, cross licensing of intellectual
property, or joint ventures, but do not wish to leak their next
steps to competitors.

There are also processes which describe the contagion of new ideas
and practices in which the credibility of information or the
willingness to adopt an innovation requires independent confirmation
from multiple sources. Unlike a `simple' biological contagious agent
carrying a disease, which can be transferred through a single
contact between two individuals, ideas and opinions (`complex'
agents) may need to be heard from multiple contacts before being
adopted~\cite{centola05}. Whether one considers teenagers deciding
to buy a new brand of jeans or farmers starting to plant a new type
of corn, the decisive event may not be hearing about an innovation,
but observing enough people participating to be convinced that the
innovation should be adopted~\cite{strang98diffusion,ryan43corn}.

The presence of closed triads enhances the probability that complex
contagion can spread on a network. Social networks tend to have a
much higher probability of closed triads than the equivalent random
networks \cite{Watts98smallworld,newman03structure}. An intuitive
reason is given by structural balance theory
\cite{Cartwright56balance} which states that ties tend to be
transitive: if a node is connected to two other nodes (is a member
of two diads), those two nodes are much more likely on average to be
connected than two randomly chosen nodes. Recently, it has also been
shown that many real world networks, including social networks,
contain overlapping k-cliques~\cite{Palla05community}. Within a
k-clique, each of the $k$ nodes is connected to each of the other
$k$ nodes, forming a densely knit community containing ${k \choose
3}$ closed triads. Two cliques were considered overlapping if they
shared $k-1$ nodes, and the question was posed whether these
overlapping cliques themselves form a network containing a fraction
of the network (the network percolates). In contrast, in this paper,
we are interested not in the overlap of cliques, but the strength of
ties between individuals. A message can be passed between two
communities, even if they share only one individual in common, as
long as that individual has strong ties within both communities.

Our results are as follows. Given the potential importance of closed
triads both in assembling varied expertise and in the diffusion of
innovation, we first determine how they are linked together in
observed social networks. We find that removing non-transitive ties
from these social networks shrinks the giant component, but does not
break it up. These results show that social networks are composed of
overlapping communities, with each community providing strong ties,
and the overlap providing a way to traverse the network using strong
ties. Secondly, we seek to  quantify the impact this local
structural requirement has on the global properties of a network,
such as the phase transition in the emergence of a giant component.
To this end, we model a random graph constructed entirely of closed
triads and compare its properties to that of an Erd\"{o}s-Renyi
graph with the same number of nodes and edges. We derive the result
that the giant connected component occurs at the same average
connectivity (average degree $\langle k \rangle = 1$), but that it
does not grow as quickly in the triad graph as the average
connectivity increases further. Numerical simulations reveal that
the average shortest path is quite similar in both networks.
Essentially, requiring transitive closure allows fewer nodes to be
connected (since 1/3 of the links must be redundant rather than
reaching out to connect additional nodes). However, the resulting
connected component will have an average shortest path that scales
logarithmically with the size of the graph, just as it would in an
Erd\"{o}s-Renyi graph.

\section{Social networks without weak ties \label{realworld}}
In order to study the connectedness of social networks without weak
ties, we analyzed two data sets. The first, and smaller, data set is
the social network of the Club Nexus online community at Stanford in
2001~\cite{adamic03social}. Much like many later online social
networking services, it allowed individuals to sign up and list
their friends on the site. The `buddy' lists were aggregated into a
single social network of reciprocated links. Within a few months of
its introduction, Club Nexus attracted over 2,000 undergraduates and
graduates, together comprising more than 10 percent of the total
student population. The Club Nexus network is only a biased subset
of the complete student social network because students had free
choice of how many friends to list. Nevertheless, the data does
provide a proxy of the true social network, from which one can
derive interesting properties. For example, triangles are quite
prevalent in this network, with a clustering coefficient of 0.17,
which is 40 times greater than what it would be for an equivalent
Erd\"{o}s-Renyi random graph. The average distance between any two
individuals is just 4 hops.

Adamic et al.~\cite{adamic03social} found that edges with high
betweenness, where betweenness reflects the number of shortest paths
that traverse the edge, tended to connect people with less similar
profiles. These profiles included information about the student's
year, field of study, personality, hobbies and other interests. The
observation that ties of high betweenness lie between dissimilar
individuals supports the hypothesis that weak ties bridge different
communities. Edges with high betweenness also tend to not be part of
closed triads, because each edge in the triad provides a possible
alternate path. In fact, a recently-devised clustering algorithm
relies on identifying communities by removing edges that participate
in fewest closed triads and longer
loops\cite{FilippoRadicchi03022004}. It is therefore a concern that
removing non-transitive ties from a network would tend to break it
apart into disconnected communities. This would mean that diverse
expertise may not be reachable and new innovations may not flow
throughout the network.

\begin{table}
\begin{tabular*}{0.90\columnwidth}%
     {@{\extracolsep{\fill}}lll}
  component size & Club Nexus  & Club Nexus \\
  & &  without weak ties \\
  \hline  
  2246  & 1 & 0\\
  1763 & 0 & 1 \\
  6 & 0 & 1 \\
  5 & 1 & 1 \\
  4 & 1 & 2 \\
  3 & 2 & 4 \\
  2 & 8 & 0 \\
  1 & 227 & 710 \\
\end{tabular*}
\caption{Distribution of connected components in online
communities.} \label{ClubNexusCCs}
\end{table}

In the case of the Club Nexus network, we can dismiss the concern,
because the network is robust with respect to the removal of weak
links, which account for 19\% of all links. Rather than breaking up
into many disconnected communities, the network sheds some nodes and
shrinks modestly. Most obviously, the 239 leaf nodes cannot be part
of triangles because they link to just one other node. They each
become a disconnected component with the removal of weak ties, which
is justified in this context because they are peripheral actors.
Table~\ref{ClubNexusCCs} shows the distribution in size of the
connected components for the original network and the network with
weak links removed. Note that both networks have a giant component
containing the majority of the nodes. The removal of weak ties does
not separate communities of large size---the largest one is composed
of just 6 nodes. The removal of weak ties does cause a slight
increase in the the average shortest path between reachable pairs.
Although the fraction of reachable pairs drops from 72\% to 51\%,
the average shortest path increases from 3.9 hops to 4.1.

The next network we consider is the network of AOL Instant Messenger
(AIM) links submitted to the website \url{buddyzoo.com}. The system
uses Buddy Lists to show users which buddies they have in common
with their friends, to visualize their Buddy List, to compute
shortest paths between screennames, and to show each user's prestige
based on the PageRank~\cite{page98pagerank} measure applied to the
network. Our anonymized snapshot of the data is from 2004 and
includes 140,181 users who submitted their buddy lists to the
BuddyZoo service, as well as 7,518,816 users who did not explicitly
register with BuddyZoo but were found on the registered users' Buddy
Lists. This is therefore a rather large social network. It was
previously studied to determine whether direct links can be
concealed in the network, for example to manipulate an online
reputation mechanism~\cite{hogg04reputation}. In the context on
BuddyZoo, this would mean that two people would remove each other
from their Buddy Lists in an attempt to hide their connection. But
unless they share no other `buddies' in common, they would still be
linked as 'friends of friends' and arguably would have a more
difficult time denying acquaintance. 9\% of the users have only a
single connection, and would disconnect themselves from the network
if they were to remove it. Of the remaining pairs of users, only
19\% could remove their direct link and be at least distance 3 from
each other, while all others would remain friends of friends. This
is equivalent to asking what percentage of the edges are parts of
triangles, which is the question we are currently interested in.
\begin{figure}[tbp]
\begin{center}

  \includegraphics[width=0.40\textwidth]{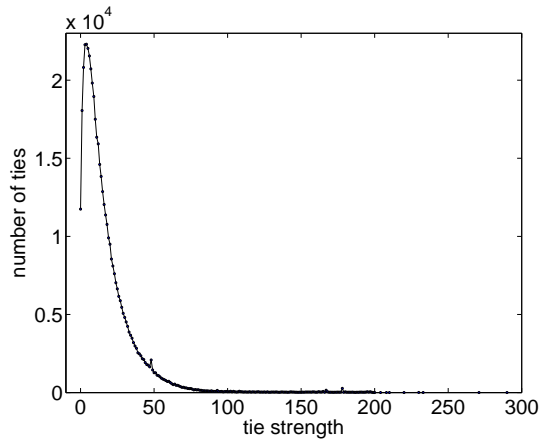}

\caption{The distribution of the strength of ties, measured as the
number of triads each tie participates in. } \label{incommonhist}
\end{center}
\end{figure}

\begin{figure}[tbp]
\begin{center}

  \includegraphics[width=0.45\textwidth]{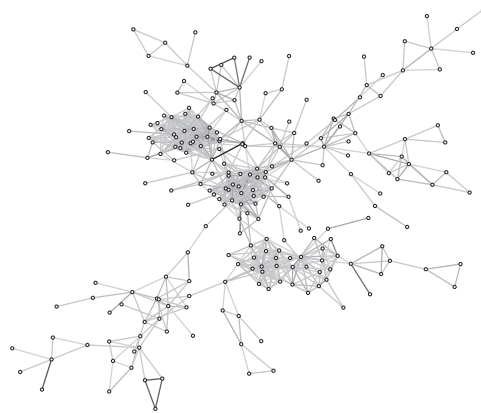}

\caption{ The largest component of the reduction of the BuddyZoo
network where each tie participates in at least 47 triads. The
triads themselves are not all shown --- only the ties that share a
threshold number of them.} \label{corenetwork}
\end{center}
\end{figure}

In order to determine the presence of strong ties, we consider only
users who explicitly registered with BuddyZoo, but we allow an edge
to be considered transitive if it is part of a closed triad that
includes an unregistered user. This is because we know that two
people share a contact, even if that contact did not register. We
exclude shared contacts that have indegree greater than 1000,
because those could be AIM bots (automated response programs). We do
not include unregistered contacts in the network itself because
their Buddy List information is incomplete. The degree distribution
is highly skewed and there are many isolates in the network. On
average, each user is connected via a reciprocated tie to 6.83 other
registered BuddyZoo users. We require a tie to be reciprocated,
since it is possible for one AIM user to add someone to their buddy
list without that person adding them in turn.

\begin{table}
\begin{tabular*}{0.90\columnwidth}%
     {@{\extracolsep{\fill}}lll}
  component size & BuddyZoo  & BuddyZoo \\
  & &  without weak ties \\
  \hline  
  124672 & 1 & 0\\
  122066 & 0 & 1\\
  21-40 & 0 & 1 \\
  11-20 & 11 & 14\\
  10 & 4 & 6 \\
  9  & 5 & 5\\
  8  & 7 & 9\\
  7  & 7 & 10\\
  6  & 15 & 16\\
  5  & 37 & 36\\
  4  & 64 & 73\\
  3 & 126 & 168\\
  2 &  591 & 685\\
  1  & 7279 & 9413\\

\end{tabular*}
\caption{Distribution of connected components in the BuddyZoo AOL
instant messenger community. A tie is considered weak if two users
who list each other on their buddy lists do not list a third person
in common.}
\end{table}

As in the case of the Club Nexus social network, we find that
removing weak ties does not have a dramatic effect on the BuddyZoo
network. Although several communities containing a couple of dozen
nodes do split off, the giant component shrinks modestly, from
occupying 88.9\% of the graph to occupying 87.5\% of it. The average
shortest path increases by a fraction of a hop from 7.1 to 7.3.
Usually any lengthening in the path decreases the probability of a
successful transmission if the probability that the message is
transferred at each step is less than 1~\cite{watts2002search}.
However, we do not observe considerable lengthening of the average
shortest path until we impose a higher threshold on tie strength.
\begin{figure}[t]
\begin{center}

  \includegraphics[width=0.45\textwidth]{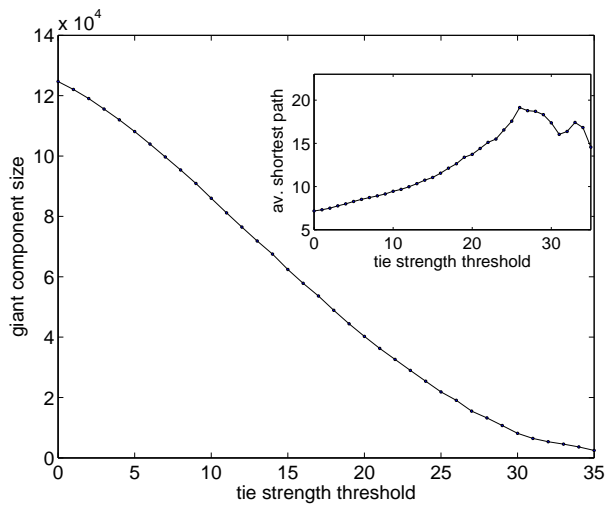}

\caption{The size of the giant component as only ties of a minimum
strength (measured in the number of triads it is a part of) are kept
in the network. The inset shows the growth of the average shortest
path between connected pairs. } \label{thresholdnetwork}
\end{center}
\end{figure}
In order to consider more restrictive requirements on tie strength,
we vary the strength threshold as follows: rather considering any
tie in a single closed triad to be strong, we require that it be
part of at least $j$ closed triads. Figure~\ref{incommonhist} shows
the distribution of tie strengths, where the mean number of shared
ties is 17.4 and the median is 13. Figure~\ref{corenetwork} shows
the largest component of nodes where each tie participates in at
least 47 triads. There are several dense cliques, but the largest
component is quite small - only 233 nodes. To investigate how
rapidly the giant component shrinks and how much the average
shortest distance changes, we consider reduced networks where only
ties of above threshold strength, measured by the number of triads
the tie participates in, are kept. Figure~\ref{thresholdnetwork}
shows the giant component size and average shortest path between all
connected pairs as the threshold is increased from zero to 35
triads. We observe that the giant component shrinks gradually,
indicating that a substantial portion of the network is spanned by
ties of moderate strength. This would indicate that the network is
composed of overlapping communities rather than separate communities
that are bridged by weak ties. What is more, removing weak ties does
not separate large communities from one another. Rather, a few
smaller communities and many isolates are spun off as the tie
strength threshold is increased. Removing weak ties has an
additional cost beyond isolating some individual nodes and smaller
communities --- it increases the average shortest path between
reachable pairs. So even though the giant component is shrinking, we
are removing the shortcuts that span it. The average shortest path
more than doubles as we increase the threshold from 1 to 25.

The strong tie robustness of the Club Nexus and BuddyZoo networks is
encouraging, especially in comparison to what one might expect in a
Watts-Strogatz (WS) type small world model~\cite{Watts98smallworld}
or an Erd\"{o}s-Renyi graph. In the WS model, the network is
constructed from a lattice where each node is connected to $k$
neighbors on each side. For $k > 1$, this means that each node
participates in local closed triads. In the model, a fraction $p$ of
the links are rewired with one endpoint placed randomly among the
nodes. It is the presence of these random links that gives the WS
model a shortest path that scales logarithmically with the size of
the graph. Such a link is unlikely to be part of triangle however,
since the probability of any two nodes linking randomly is
proportional to $1/N$ in such a graph. Therefore, removing weak
links in a WS model removes the shortcuts, leaving an average
shortest path that scales linearly with the size of the graph.
Assuming that nodes close together on the lattice share similar
information, one would need to make many hops in order to find novel
information. In section~\ref{sec:phasetransition}, we will show that
the occurrence of strong ties in an Erd\"{o}s-Renyi graph is
unlikely unless the average degree increases with the number of
nodes in the network. Therefore, removing all edges that are not
part of a triangle will isolate most of the nodes in random graphs
where the average degree is constant or nearly constant with respect
to the number of nodes.

\section{Model of a random triangle graph}

Given the results of the previous section, where we see a very high
prevalence of transitive ties and a robustness of the network with
respect to removal of weak ties, we seek to answer the basic
question of the cost of requiring all ties to be transitive. In
order to do this we consider the very simplest model of a random
graph where every edge between two nodes is part of at least one
closed triad, and investigate some properties of the graph
analytically. In essence, the graph is composed entirely of
triangles, and we model this kind of graph by assigning links among
any three randomly chosen nodes in the graph. Strictly speaking, for
a graph with $|V|=N$ nodes, there are $N \choose 3$ possible
combinations of nodes that can form a triangle. Each triangle forms
with probability $b$, so that on average we randomly choose $M = b
\times {N \choose 3}$ triplets of nodes and link them with three
edges.

Note that our method of constructing transitive graphs is similar to
a particular instance of the Newman \cite{Newman03clusterednetworks}
model for constructing highly clustered graphs. In the Newman
clustered network model, one takes a bipartite network of
individuals and groups. One then constructs a one-mode projection of
the random graph by adding, with a given probability $p$, edges
directly between individuals who belong to the same group. However,
unlike \cite{Newman03clusterednetworks}, in our model the
probability for nodes to connect to each other in the same group is
1, and the number of members in each group is constant at 3.

\subsection{Degree distribution}

We consider the degree distribution of the graph starting from the
distribution of a node belonging to $k$ closed triads.

For each node $u$, there is a total of $R={N-1 \choose 2}$ possible
triangles which have $u$ as one of the vertices. And, for each
triple of vertices, the probability of being selected to have links
in the graph is $b$. Let $r_m$ be the probability for a node belong
to $m$ chosen triples. Then

\begin{equation}
r_m={R \choose m} b^m (1-b)^{R-m}.
\end{equation}

On the other hand, we will now show that it is unlikely that our
fixed node $u$ is part of two triangles with an edge in common.  Our
node $u$ has degree $k$ if, for some $m$, node $u$ is in $m$ chosen
triples on a total of $k$ distinct nodes aside from $u$.  It is
straightforward to show that $k/2\le m\le\binom{k}{2}$.  In fact,
for even $k\ll N$, most of the probability is in the case $m=k/2$,
as we now show.  For even $k$, the probability that $u$ has degree
$k$ is the probability that $u$ is in exactly $k/2$ chosen triples,
adjusted for collisions of edges.  Collisions affect the probability
of degree $k$ in two ways---$u$ may be in exactly $m=k/2$ triples
but a collision reduces the contribution to the probability of
degree $k$, or $u$ may be in $m>k/2$ chosen triples but collisions
increase the contribution to the probability that the degree is $k$.

Conditioned on $u$ falling in exactly $m$ chosen triples, all sets
of $m$ triples are equally likely.  There are
$\binom{R}{m}=\Theta\left(\frac{N^{2m}}{2^m m!}\right)$ possible
sets of $m$ triples.  Next, we want to count the number of sets of
$m$ triples involving exactly $j$ neighbors of $u$, for $j\le 2m$.
We can pick the $j$ neighbors as a set in $\binom{N-1}{j}$ ways,
but then we need to assign roles to the $j$ neighbors based on
collision multiplicity.  For example, suppose 4 triples among five
neighbors $A,B,C,D,E$ of $u$ might be $\{u,A,B\}, \{u,A,C\},
\{u,A,D\}, \{u,B,E\}$.  We can choose $A,B,C,D,E$ as a set; pick
an element for the role of $A$ (that appears three times) in 5
ways; given that, pick an element for the role of $B$ in 4 ways;
then $E$ in 3 ways, and the remaining elements take the
interchangeable roles of $C$ and $D$, for a total of
$5\cdot4\cdot3\le 5!$ orderings).

For us, a crude bound for the orderings of roles will suffice.
There are at most $2m-j$ collisions counting multiplicities, and
so at most $2m-j$ neighbors of $u$ that can be in more than one
triple---play a non-trivial role.  There are at most $2m-j$ roles.
So the number of ways to assign non-trivial roles is at most
$(2m-j)^{2m-j}$. So the number of sets of $m$ triples involving
exactly $j$ neighbors of $u$ is at most
$\binom{N-1}{j}(2m-j)^{2m-j}$.  Thus the ratio of these to the
number of sets of $m$ disjoint triples is
\begin{eqnarray*}
\frac{\binom{N-1}{j}(2m-j)^{2m-j}}{\binom{R}{m}}
& \le & O\left(\frac{N^j(2m-j)^{2m-j}2^m m!}{j!N^{2m}}\right)\\
& \le & O\left(\frac{((2m-j)/N)^{2m-j}2^m m!}{j!}\right).
\end{eqnarray*}
We are intereseted in the case $2m-j\ge 1$.  If $m$ and $j$ are
constants, then we can ignore $2^m m!/j!$, and we get
\begin{eqnarray*}
\frac{\binom{N-1}{j}(2m-j)^{2m-j}}{\binom{R}{m}}
& \le & O\left(\frac{((2m-j)/N)^{2m-j}2^m m!}{j!}\right)\\
& \le & O\left(1/N\right).
\end{eqnarray*}

By choosing the appropriately small probability $b$ of choosing a
triple, we may assume that $m$ and $j$ are much smaller than $N$.
But we cannot necessarily assume $m$ and $j$ are constants; for
example, we may have $m!$ comparable to $N$.
We now consider the case where $j$ or $m$ grows (slowly) with $N$, and
where $N$ is sufficiently large.  If $m\le j$, then
$2^m m!/j!\le \binom{j}{m}^{-1}\le 1$.  It follows that
\begin{eqnarray*}
\frac{\binom{N-1}{j}(2m-j)^{2m-j}}{\binom{R}{m}}
& \le & O\left(\frac{((2m-j)/N)^{2m-j}2^m m!}{j!}\right)\\
& \le & O\left(((2m-j)/N)^{2m-j}\right)\\
& \le & O(N^{-1}).
\end{eqnarray*}
On the other hand, if $m>j$, then $2m-j>m$, so
\begin{eqnarray*}
\frac{\binom{N-1}{j}(2m-j)^{2m-j}}{\binom{R}{m}}
& \le & O\left(\frac{((2m-j)/N)^{2m-j}2^m m!}{j!}\right)\\
& \le & O\left(((2m-j)/N)^{2m-j}(2m)^m\right)\\
& \le & O\left((2m(2m-j)/N)^{2m-j}\right).
\end{eqnarray*}
If $2m-j=1$, this is $O(2m/N)\le N^{-1+o(1)}$.  If $2m-j>1$, then,
since we may assume that $2m\ll\sqrt{N}$, we have
\begin{eqnarray*}
\frac{\binom{N-1}{j}(2m-j)^{2m-j}}{\binom{R}{m}}
& \le & O\left((2m(2m-j)/N)^{2m-j}\right)\\
& \le & O\left(((2m-j)/\sqrt{N})^{2m-j}\right)\\
& \le & O\left(((2m-j)^2/N)^{(2m-j)/2}\right).
\end{eqnarray*}
This is $O\left((2m-j)^2/N\right)\le N^{-1+o(1)}$.

We conclude that the effect of collisions is small in any case.  Thus
we get the probability of $u$ having degree $k$ is
\begin{equation}
p_k = \left \{
     \begin{array}{ l l }
     {R \choose {\frac{k}{2}}} b^{\frac{k}{2}}
     (1-b)^{R-\frac{k}{2}}\pm N^{-1+o(1)}
     & \mbox {if $k$ is even} \\
     N^{-1+o(1)},
     & \mbox{if $k$ is odd}\\ \end{array} \right.
\end{equation}

After ignoring the additive amount $\pm N^{-1+o(1)}$,
the corresponding generating function is given by
\begin{equation}
G_0(z) = \sum^R_{k=0}{R \choose k} b^k
(1-b)^{R-k}z^{2k}=[bz^2+1-b]^R
\end{equation}

The average degree $\langle k \rangle$ is then given by:
\begin{equation} \label{4}
\langle k \rangle = G_0^\prime(1) = b(N-1)(N-2)
\end{equation}

And thus, we have the relationship between average degree $\langle k
\rangle$ and the probability of any three nodes being connected by a
triangle $b$:

\begin{equation}
b=\frac{\langle k \rangle}{(N-1)(N-2)}
\end{equation}

When $\langle k \rangle =O(1)$, $b=O\left( \frac{1}{N^2} \right)$.

\subsection{Accidental triangles and the clustering coefficient}

We should notice that in our model, the expected number of triangles
in the network is not exactly $b \times {N \choose 3}$. There is the
possibility of forming an ``accidental" triangle, which can occur
when the pairs of nodes $a$ and $b$, $b$ and $c$, and $a$ and $c$
are linked, but the triangle ${a,b,c}$ was not among the $b \times
{N \choose 3}$ initially chosen triangles. The probability $b\prime$
of this occurring is the probability that no triangle was
intentionally formed between the $a$, $b$, and $c$: $1-b$ times the
probability that each of the three edges does occur in a triangle
other than ${a,b,c}$.

\begin{equation}\label{5}
b'=(1-b)[1-(1-b)^{(N-3)}]^3
\end{equation}

In this way, we know that the total expected number of triangles in
this graph is $a\times{N \choose 3}$, where $a = b+ b'$.

Thus, the ratio between the actual number of triangles in the graph
and the input number of triangles is:

\begin{equation}\label{6}
\Delta = \frac{a}{b} = 1+\frac{(1-b)[1-(1-b)^{(N-3)}]^3}{b}
\end{equation}

However, $b'$ is very small compared with $b$, when the average
degree of a node in the graph is a constant independent of the
growth of the total number of nodes $N$. Since we have shown that
$b=O(\frac{1}{N^2})$, then it is not hard to see that the ratio of
the probability for any three nodes to be part of an accidental
triangle and the probability for them to be a triangle that is
constructed by randomly choosing groups is:
\begin{equation}\label{7}
\frac{b'}{b} = \frac{(1-b)[1-(1-b)^{(N-3)}]^3}{b}= O \left(
\frac{1}{N} \right)
\end{equation}

\begin{figure}[tbhp]
\begin{center}
  \includegraphics[width=0.9\columnwidth]{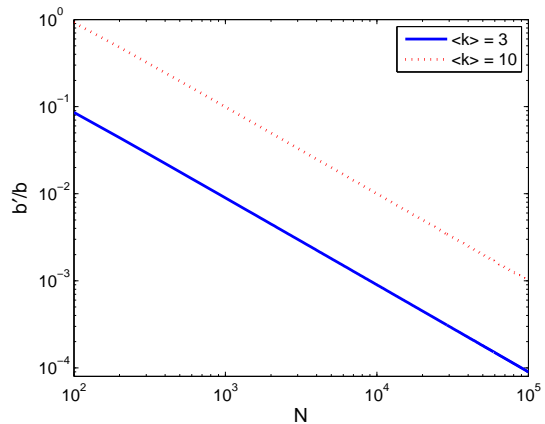}
\caption{The ratio of the number of accidentally formed triangles to
the number randomly chosen by the model. For fixed average degree
and increasing number of nodes, the ratio of accidentally formed
triangles drops as $1/N$.} \label{fig:accidental}
\end{center}
\end{figure}

Thus, we can see that when $N$ is large, and the average degree
$\langle k \rangle$ is independent of $N$, then the chance of
forming an accidental triangle is quite small compared to the
triangles randomly drawn in constructing the model.
Figure~\ref{fig:accidental} shows the relation between $b'$ and
average degree $\langle k \rangle$.

\begin{figure*}[t]
\begin{center}
\begin{tabular*}{\textwidth}{@{\extracolsep{\fill}}lcr}
  \includegraphics[width=0.30\textwidth]{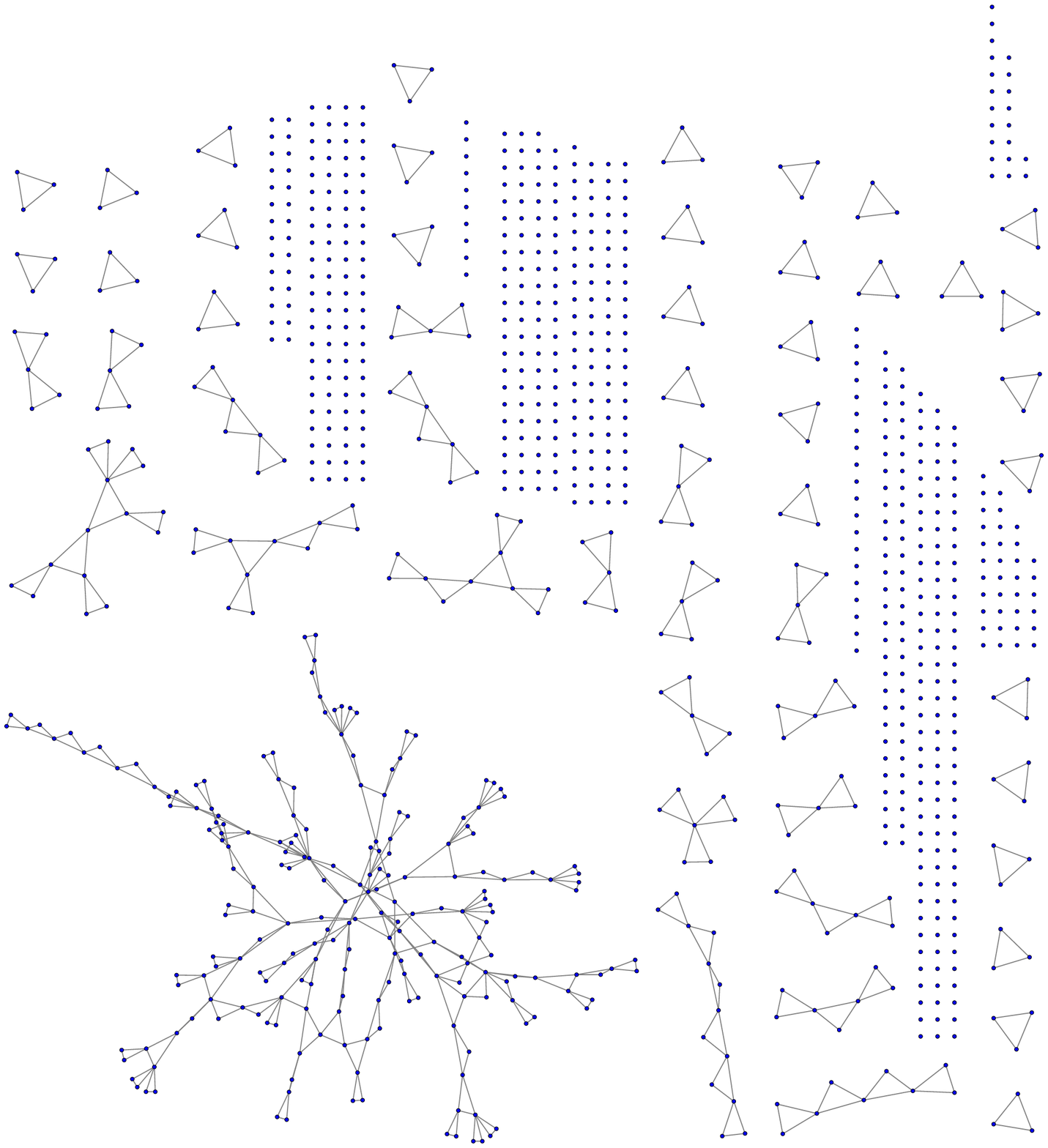} &
  \includegraphics[width=0.30\textwidth]{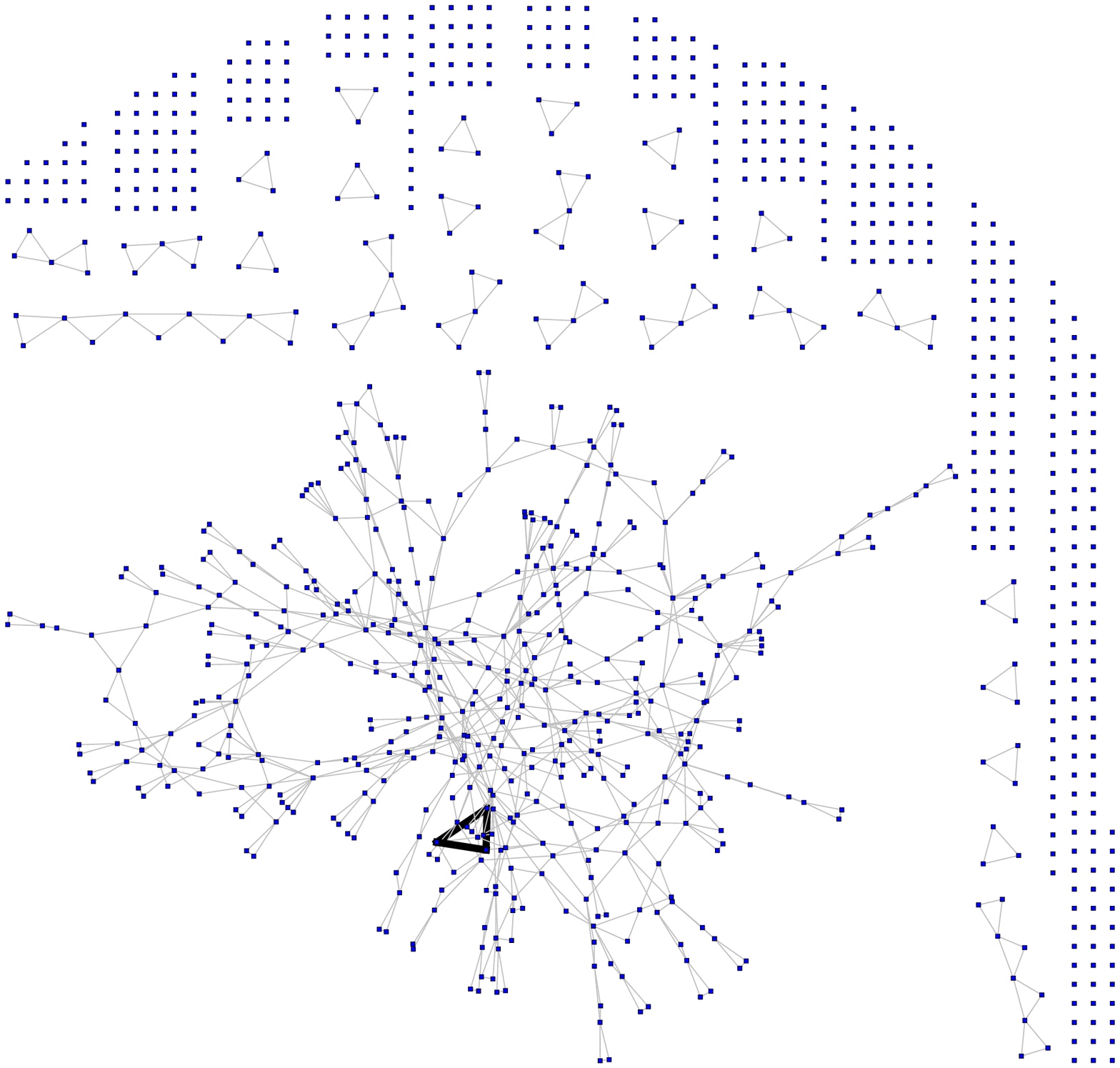} &
  \includegraphics[angle=90,width=0.28\textwidth]{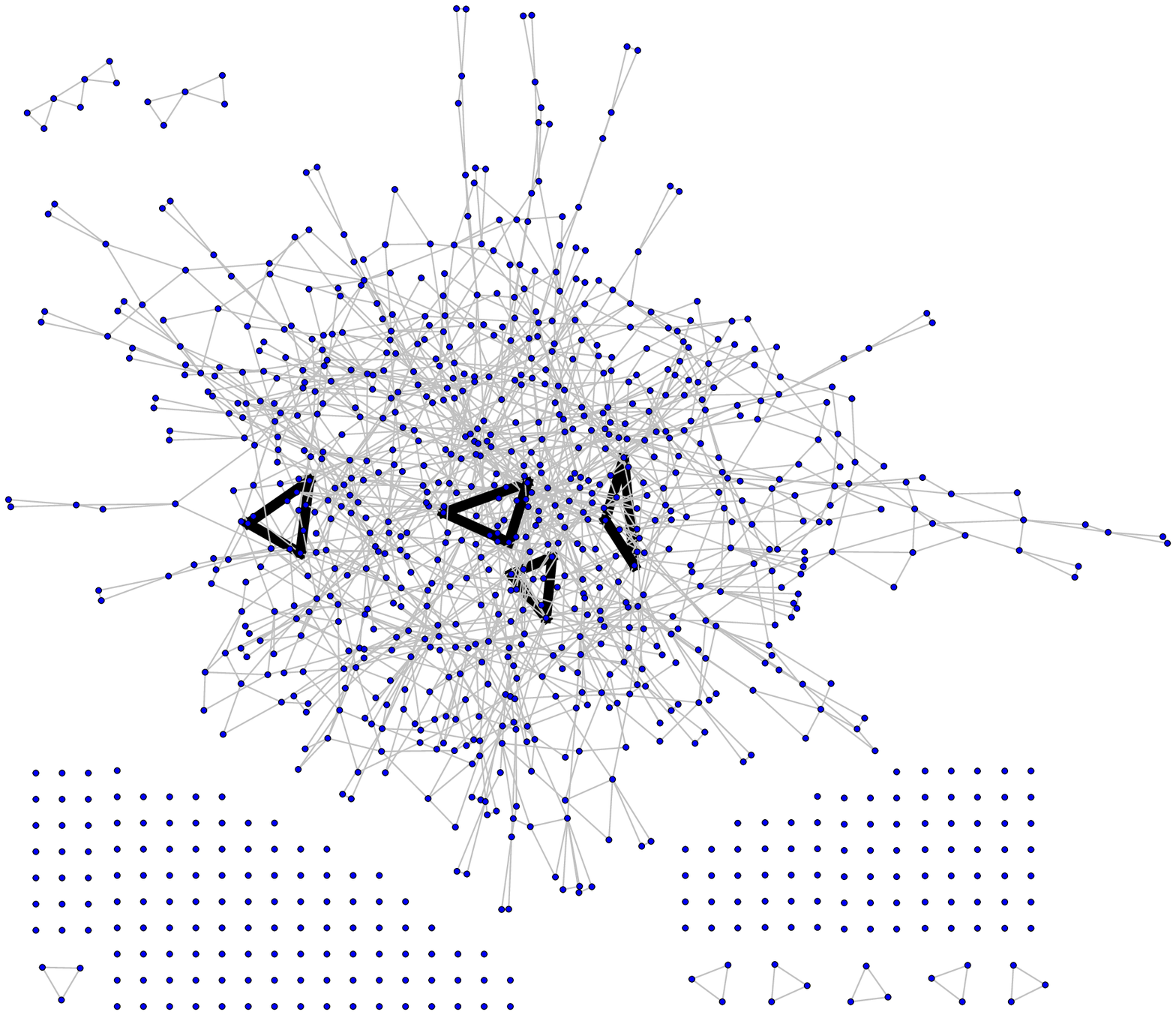} \\
  (a) $N = 1000$, $M = 200$ & (b) $N = 1000$, $M = 300$ & (c) $N = 1000$, $M = 500$\\
\end{tabular*}
\caption{Examples of triangle graphs with 1000 nodes with varying
numbers of triangles $M$. Accidental triangles are marked with bold
lines.} \label{fig:visualizations}
\end{center}
\end{figure*}

In Figure~\ref{fig:visualizations} we show three instances of a
randomly generated graph of triangles. Each graph has $1,000$ nodes,
but we form different numbers of triangles. Even though a giant
component exists for each graph, it is only once the number of
triangles equals the number of nodes that we observe a few random
triangles forming. Therefore the formation of accidental triangles
does not have a substantial effect on the derivations below.

The clustering coefficient $C$ is a measure of the prevalence of
closed triads in a network \cite{Watts98smallworld,
newman03structure}. The expectation of the total number of connected
triples of nodes (open and closed triads) in the graph is $
N_{triple} = N \times \sum_k {k \choose 2} p_k$, and the number of
closed triads is $N_{\Delta}\approx b \times {N} {N \choose 3}$
since the number of accidental triangles is small. Thus the
clustering coefficient is:
\begin{eqnarray*}
C &=& \frac{3N_{\Delta}}{N_{triple}}    \\
  &\approx& \frac{3b {N \choose 3}}{N \times \sum_k {k \choose 2} p_k}  \\
  &=& O(1)
\end{eqnarray*}

We can see that when $N$ is large, the clustering coefficient of
our graph is:
\begin{equation} \label{8}
C = O(1)
\end{equation}
which is significantly larger than the $O(N^{-1})$ clustering
coefficient in an Erd\"{o}s-Renyi Random graph. For many types of
real world networks, it has been shown that $C=O(1)$
\cite{newman03structure}, so it is of interest to see how removing
weak ties in real networks changes the clustering coefficients.

\subsection{Phase transition and the giant component\label{sec:phasetransition}}

For the derivation of the phase transition and size of giant
component, we loosely follow the generating function methods for
clustered graphs in~\cite{newman03structure}. The phase transition
is also known as the percolation threshold - the average degree at
which a finite fraction of the network is connected, forming a giant
component. In Part A, we have given $r_m$, the probability for a
node belong to $m$ triangles. Thus, averaging over all individuals
and triangles, we have the mean number of triangles a node belongs
to: $\mu = \sum_m mr_m$.

The probability of having two edges within the triangle is 1, and
the probability of having any other number is 0. Therefore, the
generating function of the number of edges for each node within a
triangle is

\begin{equation}\label{9}
  h(z) = z^2
\end{equation}

Furthermore, for a node $A$ in the graph, the total number of other
nodes in the whole graph that it is connected to by virtue of
belonging to triangles is generated by:
\begin{equation}\label{10}
G_0(z)=\sum^\infty_{m=0}r_m(h(z))^m
\end{equation}
where $r_m$ is the probability for a node to belong to $m$ groups as
we defined before. This is also the generating function of the
distribution of the number of nodes one step away from node $A$.

The generating function of the distribution of the number of nodes
two steps away from $A$ is $G_0(G_1(z))$, where $G_1(z)$ is the
generating function for the distribution of the number of neighbors
of a node arrived at by following an edge (excluding the edge that
was used to arrive at the node):

\begin{equation}\label{11}
G_1(z)=\mu^{-1}\sum^\infty_{m=0}mr_m(h(z))^{m-1}
\end{equation}

The necessary and sufficient condition for a giant component to
exist, is when, averaging over all the nodes in the graph, the
number of nodes two steps away exceeds the number of nodes one step
away~\cite{newman-2001-64}, which can be expressed as:
\begin{equation}\label{12}
[\partial_z(G_0(G_1(z))-G_0(z))]_{z=1}>0
\end{equation}

Thus, we get the condition for the existence of a giant component
in this graph:

\begin{eqnarray*}
  ((\mu^{-1} \sum^{\infty}_{m=0}m(m-1)r_mz^{m-2}) \cdot
  h'(z))|_{z=1} &>& 1    \\
  2\mu^{-1} \sum^{\infty}_{m=0}m(m-1)r_m &>& 1 \\
  \frac{R(R-1)b}{Rb} &>& \frac{1}{2}
\end{eqnarray*}
After simplifying the above equation, the condition is:
\begin{equation}\label{13}
b > \frac{1}{N^2-3N}
\end{equation}

Since we will compare this graph with an Erd\"{o}s-Renyi random
graph with the same average degree $\langle k \rangle$, we express
the condition for the existence of giant component in terms of the
average degree given by Equation~\ref{4}:

\begin{equation}\label{14}
\langle k \rangle  >  1+ \frac{2}{N^2-3N}
\end{equation}

As $N\rightarrow \infty$, the condition is $\langle k \rangle  >
1$. An interesting point is that this is exactly where the phase
transition occurs in an Erd\"{o}s-Renyi graph. Therefore, the
requirement that all edges be transitive does not delay the
appearance of the giant component. It does however have a
tempering effect on the rate of growth of the giant component as
we will see below.

When a giant component exists in the graph and the probability for
a node to whom $A$ is connected to not belong to it is $s$, the
size of the giant component is given by:
\begin{eqnarray}\label{15}
S &=& 1-G_0(s_0)    \\
  &=& 1-\sum^\infty_{m=0}r_m(s_0^2)^m   \\
  &=& 1-(bs_0^2 +1 -b)^R
\end{eqnarray}
where $s_0$ is the solution of the function:
\begin{eqnarray}\label{18}
s &=& G_1(s)    \\
 &=& \mu^{-1}\sum^\infty_{m=0}mr_m(s^2)^{m-1}  \\
 &=& (bs^2+1-b)^{R-1}
\end{eqnarray}

As we have assumed $S > 0$, we know that $s$ must be some value
larger than 0 and smaller than 1, and thus $s=1$ is a trivial
solution of the function.

\begin{figure}[tbhp]
\begin{center}
  \includegraphics[width=0.9\columnwidth]{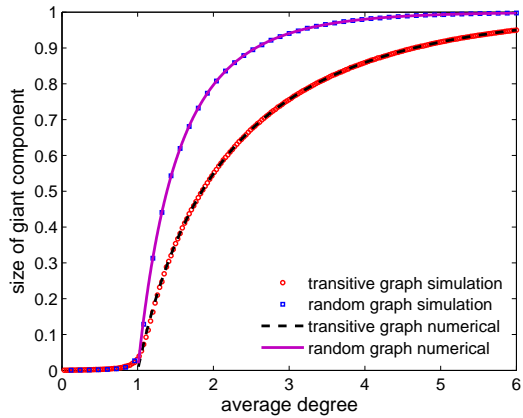}
\caption{Comparison of numerical simulations with analytical
solutions for the fraction of the network occupied by the giant
component of a 10,000 node triangle graph and the corresponding
Erd\"{o}s-Renyi graph} \label{fig:giantcomponent}
\end{center}
\end{figure}

We compare the solution $s_0$ to numerical simulations of networks
of random triangles. Each network contains $N=10,000$ nodes, and
we select $M$ random triangles to connect from the $N$ nodes. For
each value of $M$ we generate 50 random networks and average the
size of the giant component. The results, shown in
Figure~\ref{fig:giantcomponent} show excellent agreement between
the analytical prediction and the numerical simulation. For
comparison, we show both the numerical prediction and analytical
result for the size of the giant component in an Erd\"{o}s-Renyi
random graph with the same number of nodes and edges. The size of
the giant component in the Erd\"{o}s-Renyi graph is given by the
solution $s$ to the equation $s = 1 - \exp(-\langle k \rangle s)$.
>From the figure, we can see that as average degree grows, the
phase transitions of the transitive graph and the random graph
occur at the same time, while the size of giant component of the
Erd\"{o}s-Renyi graph grows more quickly as we increase the
average degree. An intuitive explanation is that in an
Erd\"{o}s-Renyi graph one need not expend a `closure' edge to
close a triad. Rather, that edge can be used to connect a
disconnected node or small component to the giant component.

The fact that the phase transition occurs at the same average degree
for both the Erd\"{o}s-Renyi and transitive network shows that the
requirement of transitivity does not result in a need for increased
average connectivity in order for the giant component to form. Note
that the phase transition in our model, where all edges are the
result of the addition of triangles, is quite different from what it
is in a graph that would result from taking a simple Erd\"{o}s-Renyi
graph and removing all edges that do not fall within a triangle. In
the Erd\"{o}s-Renyi graph with non-transitive edge removal the
percolation threshold occurs at a degree that scales as
$N^{\frac{1}{3}}$.

This condition for the giant component in an Erd\"{o}s-Renyi graph
with weak ties removed can be derived as follows. A giant component
of strong tries forms when, after arriving at an arbitrary triangle
$T$, the expected value of the number other adjacent triangles that
one could``move to" is equal to 1. The probability that there is a
triangle $T'$ adjacent to $T$ that is not the triangle from which we
reached $T$ is given by $2 {{N-5} \choose 2} p^3$.  There are
${{N-5} \choose 2}$ choices for the vertices in $T'$ not shared with
$T$, and two choices of the vertex shared by $T$ and $T'$ (excluding
the vertex of $T$ that is shared with the triangle we arrived from).
$p = \langle k \rangle/N$ is the probability that any two vertices
in an Erd\"{o}s-Renyi graph share an edge. Thus when $N$ is large,
the average degree at the phase transition is $\langle k
\rangle=N^{1/3}$. In several real world networks the average degree
was found to vary as $N^{\beta}$ where
$0\leq\beta\leq0.3$~\cite{leskovec05densification}. But in a random
network, this density falls short of the $N^{1/3}$ necessary to make
the accidental occurrence of closed triads (and therefore strong
ties) high enough for the network to percolate.

If one further requires that the triangles overlap not just in one
node but in two, as in the percolation of
k-cliques~\cite{derenyi01cliquepercolation}, the phase transition
occurs at a critical average degree that grows as
$N^{\frac{k-2}{k-1}}$, with $k=3$. This means that the average
degree has to grow in linear proportion to $N$ in order for a
giant component to form. Together, these two results show that the
Erd\"{o}s-Renyi random graph typically does not contain
sufficiently numerous strong ties to percolate. But as we have
shown in section~\ref{realworld}, real world social networks do
contain many strong ties that percolate. This can be intuitively
explained by the observation that new social ties typically form
in the context of geographical and sociocultural
settings~\cite{watts2002search}. In these contexts it is natural
that the ties tend to form closed triads rather than being added
independently, as they are in Erd\"{o}s-Renyi random graphs.

\section{Average Shortest Path}
Exact results for the average shortest path are difficult to
derive even for a random graph. We therefore used numerical
simulations to measure the average shortest path between all
reachable nodes as we increase the size of the network. We
selected a value of the average node degree where the giant
component existed, but did not take up all of the graph. At our
chosen value, $M=0.5 N$, there are twice as many triangles as
nodes. This constant proportion of triangles to nodes means that
$b$, the probability of any triple of nodes being connected, falls
as $1/N^2$.

\begin{figure}[tbhp]
\begin{center}
  \includegraphics[width=0.9\columnwidth]{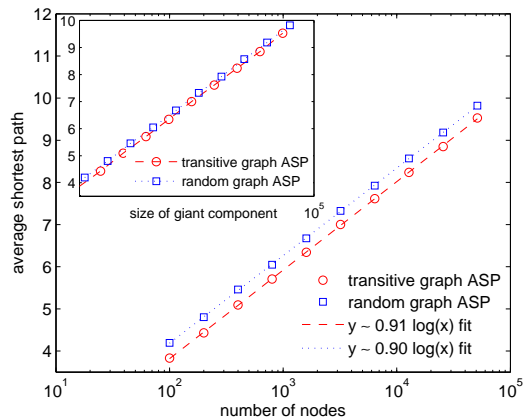}
\caption{Numerical comparison of the average shortest path in
triangle graphs and Erd\"{o}s-Renyi graphs with the same number of
nodes and edges. The inset shows the average shortest path as a
function of the size of the giant component rather than the total
number of nodes. } \label{fig:avshortpath}
\end{center}
\end{figure}

At $M = 0.5 N$, the giant component occupies 76\% of the nodes,
while in the equivalent random graph it takes up 94\% of the nodes.
This makes it difficult to directly compare the two networks, since
the average shortest path is measured between reachable pairs, and
the Erd\"{o}s-Renyi graph has more of them.
Figure~\ref{fig:avshortpath} shows that the average shortest path is
actually shorter in the triangle graph. This may be explained by the
fact that there are fewer nodes in the giant component but a greater
density of links. Once we consider the average shortest path
relative to the size of the giant component, the curves become
nearly identical for both networks. This shows that the requirement
of triadic closure does not negatively impact the average shortest
path for reachable pairs, but those pairs are fewer in number.

\section{Conclusions and future work}
In this paper we study the connectivity of strong ties in networks,
where strong ties are defined as belonging to closed triads. We find
that two real world social networks are robust with respect to
removal of weak links, in the sense that there remains a giant
component that is smaller but still occupies a majority of the
graph. We also find empirically that the removal of weak links
lengthens the average shortest path modestly. In comparison, the
removal of weak links in an WS small world network or an
Erd\"{o}s-Renyi graph would isolate the vast majority of nodes. It
is the high clustering of social networks that allows them to
transmit or gather information via strong ties.

We also pose a basic question, which is the cost paid for the
requirement of transitive ties in terms of the size of the giant
component and the length of the average shortest path. We consider
the simplest random graph model consisting entirely of closed triads
and compare it to a network where the links are randomly rewired. We
find that the giant component occurs at the same point---when the
average node degree equals 1. However, past the phase transition,
the giant component in the graph of closed triads grows more slowly
than it does in the random network. We further examine the
dependence of the average shortest path with the size of the network
and find it to be almost identical for reachable pairs in both the
triangle graph and the equivalent random network.

An unanswered question is whether more sophisticated models of
social structure~\cite{kleinberg2000navigation,
kleinberg01dynamics,watts2002search} capture the phenomenon of
strong ties that can be linked together to span an entire network.
In particular, in future work we are interested in examining the
strong tie properties of social networks where the edge
probabilities depend on the hierarchial organization of underlying
social dimensions.

\begin{acknowledgments}
We would like to thank Mark Newman and Scott Page for insightful
discussions and suggestions.
\end{acknowledgments}

\bibliography{triads}
\bibliographystyle{plain}

\end{document}